\documentclass[aps,prd,floatfix,superscriptaddress,twocolumn,showpacs]{revtex4}

\usepackage{epsfig}
\usepackage{amsmath}
\usepackage{graphicx,psfrag}
\usepackage{dcolumn}
\usepackage{bm}
\usepackage{slashed}

\newcommand{\beq}{\begin{equation}}
\newcommand{\eeq}{\end{equation}}
\newcommand{\bea}{\begin{eqnarray}}
\newcommand{\eea}{\end{eqnarray}}
\newcommand{\hf} {\frac{1}{2}}

\newcommand{\nn}{\nonumber\\}

\newcommand\fig[1]     {Fig.\,{\ref{#1}}}

\def\Tr{{\rm Tr}}

\def\eq#1{(\ref{#1})}
\def\s0#1#2{\mbox{\small{$ \frac{#1}{#2} $}}}
\def\0#1#2{\frac{#1}{#2}}

\def\mr#1{{\mathrm{#1}}}

\begin{document}

\title{Spontaneous symmetry breaking 
and optimization of functional renormalization group}

\author{I. N\'andori}
\affiliation{MTA-DE Particle Physics Research Group, P.O.Box 51, H-4001 Debrecen, Hungary}
\affiliation{MTA Atomki, P.O. Box 51, H-4001 Debrecen, Hungary} 

\author{I. G. M\'ari\'an}
\affiliation{University of Debrecen, P.O.Box 105, H-4010 Debrecen, Hungary}

\author{V. Bacs\'o}
\affiliation{University of Debrecen, P.O.Box 105, H-4010 Debrecen, Hungary}

\begin{abstract} 
The requirement for the absence of spontaneous symmetry breaking in the $d=1$ 
dimension has been used to optimize the regulator dependence of functional 
renormalization group equations in the framework of the sine-Gordon scalar 
field theory. Results obtained by the optimization of this kind were compared 
to those of the Litim-Pawlowski and the principle of minimal sensitivity optimization
scenarios. The optimal parameters of the compactly supported smooth (CSS) 
regulator, which recovers all major types of regulators in appropriate limits, have 
been determined beyond the local potential approximation, and the Litim limit of the 
CSS was found to be the optimal choice.
\end{abstract}


\maketitle

\section{Introduction}
Spontaneous symmetry breaking plays an important role in high energy 
physics or more generally in quantum field theory; as an example one has 
to mention the mass generation by the Higgs mechanism. However, in the
(0+1) dimension as a consequence of the equivalence between quantum 
field theory and quantum mechanics, a symmetry cannot be broken 
spontaneously due to tunneling \cite{zj,d1_anharmonic}. Thus, 
for one-dimensional quantum field theoretic models, the spontaneously
broken phase should vanish if their phase structures have been 
determined without using approximations.

Renormalization has relevance in quantum field theory, too, since this 
procedure is required to obtain measurable physical quantities. It can 
be performed nonperturbatively by means of the functional renormalization 
group (RG) method \cite{WP,We1993,Mo1994,internal} which was applied
successfully in many cases; let us mention quantum Einstein gravity \cite{qeg}
as a recent example. The functional RG equation for scalar fields \cite{We1993}
\beq
\label{erg}
k \partial_k \Gamma_k [\varphi] = \hf  \Tr \left[
(k\partial_k R_k) / (\Gamma_k^{(2)}[\varphi] + R_k)
\right]
\eeq
is derived for the blocked effective action $\Gamma_k$, which interpolates 
between the bare $\Gamma_{k\to \Lambda} = S$ and the full quantum 
effective action $\Gamma_{k\to 0}=\Gamma$ where $k$ is the running 
momentum scale. The second functional derivative of the blocked action 
is represented by $\Gamma_k^{(2)}$, and the trace Tr stands for the 
momentum integration. $R_k$ is the regulator function where $R_k(p\to 0)>0$, 
$R_{k\to 0}( p)=0$ and $R_{k\to \Lambda}( p)=\infty$. To solve the 
RG equation \eq{erg} one of the commonly used systematic approximations 
is the truncated derivative (i.e., gradient) expansion where $\Gamma_k$ is 
expanded in powers of the derivative of the field,
\beq
\label{deriv}
\Gamma_k [\varphi] = \int d^d x  \left[V_k(\varphi) 
+ Z_k(\varphi) \hf (\partial_{\mu} \varphi)^2 + ... \right].  
\eeq 
Further approximations such as the Taylor or Fourier expansion of 
$V_k(\varphi)$, or $Z_k(\varphi)$, are usually also applied. However, the 
usage of approximations generates two problems: (i) for the $d=1$ dimension 
the spontaneously broken phase does not vanish in the approximated RG 
flow, and (ii) physical results obtained by the approximated RG flow become 
regulator dependent (i.e. renormalization scheme dependent). Therefore, 
it is of great importance to consider how the approximations used influence 
the phase structure of one-dimensional models and the comparison 
of results obtained by various types of regulator 
functions \cite{opt_rg,litim_o(n),opt_func,Ro2010,Mo2005,qed2,scheme,
scheme_sg,minimal_sens,css,css_pms} is also a general issue. 

To optimize the scheme dependence, the Litim-Pawlowski
optimization method has been worked out \cite{opt_rg,opt_func} based on 
the convergence of the truncated flow that is expanded in powers of the field variable. 
Its advantage is that in the leading order of the gradient expansion, i.e., in the local 
potential approximation (LPA), it is possible to find the optimal choice for the parameters 
of all the regulator functions. Furthermore the Litim's optimized regulator was 
constructed, which is expected to provide us with findings closest to "the best known" 
results in LPA, e.g. critical exponents of the $O(N)$ scalar theory in 
$d=3$ dimensions \cite{litim_o(n),minimal_sens,scheme,IR}. Its disadvantage 
is that Litim's regulator is in conflict with the derivative expansion since it is not
a differentiable function. 

Another scenario for optimization through the principle of minimal sensitivity 
(PMS) was also considered \cite{minimal_sens}. Its advantage is that it can 
be applied at any order of the derivative expansions for any dimensions; i.e. 
it is possible to find the optimal choice for the parameters of a particular regulator. 
Its disadvantage is that one cannot determine the best regulator function 
among the usual exponential \cite{We1993}, power-law \cite{Mo1994} and 
Litim \cite{opt_rg} regulators through the PMS. However, the combination of the 
PMS method and the so-called compactly supported smooth (CSS) regulator 
\cite{css} provides the tool for optimization where various types of regulators 
can be directly compared to each other because the CSS recovers the major types 
of regulators in appropriate limits. This strategy has been successfully 
applied in LPA \cite{css_pms} in the framework of the $O(N)$ scalar theory in 
$d=3$ dimensions and the two-dimensional bosonized quantum electrodynamics 
(QED$_2$). In LPA the Litim regulator was found to be the most favorable 
regulator.

Our goal in this work is to open a new platform to optimize RG equations that 
represents a suitable optimization scenario beyond LPA. In this new strategy, the 
requirement of the absence of the broken phase in the case of the nonapproximated 
RG flow in the $d=1$ dimension is used to optimize the RG scheme dependence of the 
approximated one. Its advantage is that regulators can be compared to each other 
at any order of the derivative expansion for $d=1$. It is performed in 
the framework of the sine-Gordon (SG) model \cite{cg_sg} which does not require 
field-dependent wave-function renormalization; thus, the determination of RG equations 
beyond LPA is simpler than in the case of other models. On the contrary, for the $O(N)$ 
scalar theory the field-dependent wave function renormalization cannot be avoided. 
Nevertheless, the new optimization method proposed here can be applied to the 
$O(N)$ scalar theory in $d=1$, too. After testing the optimization for the power-law 
regulator \cite{Mo1994}, we optimize the CSS regulator \cite{css}. A similar 
compactly supported smooth function has been used in nuclear physics \cite{sv} 
and the connection to the CSS regulator was shown \cite{css}.

\section{Regulator functions}
Regulator functions have already been discussed in the literature by 
introducing its dimensionless form
\beq
R_k( p) = p^2 r(y),
\hskip 0.5cm
y=p^2/k^2,
\eeq
where $r(y)$ is dimensionless. For example, the CSS regulator 
introduced recently in Ref. \cite{css} is defined as
\beq
\label{css_gen}
r_{\mr{css}}^{\mr{gen}}(y) = 
\frac{\exp[c y_0^{b}/(f-h y_0^{b})] -1}{\exp[c y^{b}/(f -h y^{b})] -1}  
\theta(f-h y^b),
\eeq 
with the Heaviside step function $\theta(y)$. Let us note, that the number 
of free parameters in \eq{css_gen} can be reduced by setting $f=1$ 
without loss of generality. The CSS regulator has the property \cite{css} to 
recover all major types of regulators: the Litim \cite{opt_rg}, the power-law 
\cite{Mo1994} and the exponential \cite{We1993} ones. By choosing a 
particular normalization (i.e. fixing $y_0$) the CSS regulator reads
\begin{align}
\label{css_norm}
r_{\mr{css}}^{\mr{norm}}(y) =& \;
\frac{\exp[\ln(2) c]-1}{\exp\left[\frac{\ln(2) c y^{b}}{1 -h y^{b}}\right] -1}  
\theta(1-h y^b), 
\end{align}
where the limits are
\begin{subequations}
\label{css_norm_limits}
\begin{align}
\label{opt_lim}
\lim_{c\to 0,h\to 1} r_{\mr{css}}^{\mr{norm}} = & \;
\left(\frac{1}{y^b} -1\right) \theta(1-y), \\
\label{pow_lim}
\lim_{c\to 0, h \to 0} r_{\mr{css}}^{\mr{norm}} = & \;
\frac{1}{y^b}, \\ 
\label{exp_lim}
\lim_{c \to 1, h \to 0} r_{\mr{css}}^{\mr{norm}} = & \;
\frac{1}{\exp[\ln(2) y^b]-1}.
\end{align}
\end{subequations}
The advantage of this type of normalization is that the form~\eq{css_norm} 
reproduces all the major types of regulators with optimal parameters, 
i.e. the Litim \eq{opt_lim} with $b=1$, the power-law \eq{pow_lim} with $b=2$,
and the exponential \eq{exp_lim} with $b=1.44$. The optimal choices for the 
parameter $b$ are based on the Litim-Pawlowski optimization scenario.

\section{SG model for dimensions $1\leq d \leq 2$}
To perform the RG study of the SG model \cite{sg} beyond 
LPA it is convenient to introduce a dimensionless variable 
$\tilde\varphi = k^{(2-d)/2} \varphi$, and then the effective action reads 
\bea
\label{eaans_dimless}
\Gamma_{k} =
\int d^d x  \left[\hf z_k (\partial_\mu{\tilde\varphi})^2 
+ u_k \cos(\tilde\varphi) \right],
\eea
where $u_k$ is the dimensionful coupling of the periodic self-interaction, 
$z_k$ stands for the field-independent wave-function renormalization 
that has a dimension of $k^{d-2}$ \cite{cg_sg}. Although RG transformations 
generate higher harmonics, we use the ansatz \eq{eaans_dimless} that 
contains a single Fourier mode since in the case of the SG model it was found to 
be an appropriate approximation \cite{cg_sg}. The RG flow equations for the 
couplings of \eq{eaans_dimless} can be derived from \eq{erg} 
\bea
\label{exact_u}
k\partial_k u_k =
\int _p \frac{k\partial_k R_k}{k^{2-d} u_k}
\left(\frac{P-\sqrt{P^2-(k^{2-d} u_k)^2}}{\sqrt{P^2-(k^{2-d} u_k)^2}}\right),\\
\label{exact_z}
k\partial_k z_k = \int_p \frac{k\partial_k R_k}{2}
\biggl[
\frac{-(k^{2-d}u_k)^2P(\partial_{p^2}P+\frac{2}{d}p^2\partial_{p^2}^2P)}
{[P^2-(k^{2-d}u_k)^2]^{5/2}}\nn
+\frac{(k^{2-d}u_k)^2 p^2 (\partial_{p^2}P)^2(4P^2+(k^{2-d}u_k)^2)}
{d \, [P^2-(k^{2-d} u_k)^2]^{7/2}}
\biggr] ,
\eea
where $P = z_k k^{2-d} p^2+R_k$ and the momentum integral 
$\int_p = \int dp \, p^{d-1} \Omega_d/(2\pi)^d$ is usually performed 
numerically with the $d$-dimensional solid angle $\Omega_d$. The RG study 
of SG type models \cite{sg,qed2,qed_qcd} does not require field-dependent 
wave-function renormalization. We use normalized dimensionless parameters
$\bar z_k \equiv (8\pi) \tilde z_k $ and $\bar u_k \equiv \tilde u_k k^2/ {\bar k}$
where $\tilde z_k = k^{2-d} z_k$ and $\tilde u_k = k^{-d} u_k$ are the 
conventional dimensionless couplings and $\bar{k} = \min_{p^2} P$.

\begin{figure}[ht] 
\begin{center} 
\epsfig{file=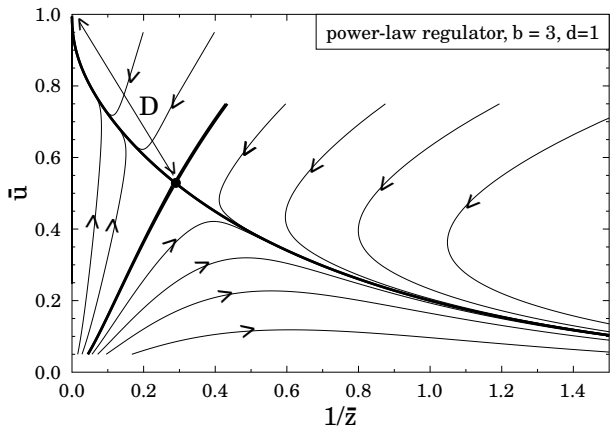,width=8.0 cm}
\caption{
\label{fig1}
Phase diagram of the SG model for $d=1$ dimensions obtained by the
numerical solution of Eqs.\eq{exact_u}, and \eq{exact_z} using the power-law 
regulator \eq{pow_lim} with $b=3$. Arrows indicate the direction of the flow. 
The distance $D$ is defined by \eq{dist}.
} 
\end{center}
\end{figure}

In $d=2$ dimensions the SG model undergoes a topological phase 
transition \cite{sg} where the critical value that separates the phases of 
the model, $1/\bar z_{\star} = 1$, was found to be independent of the choice 
of the regulator function \cite{scheme_sg}. For the $d=1$ dimension, based on 
the approximated RG flow, a saddle point $\bar u_{\star}$, $1/\bar z_{\star}$
appears in the RG flow \cite{cg_sg}; see, for example the results \fig{fig1} 
obtained by the power-law regulator \eq{pow_lim}, and thus the SG model has two 
phases. 
%
%
In fractal dimensions, $1<d<2$ the nontrivial saddle point appears in the RG 
flow, too. However, there is an important difference between the cases of fractal 
dimensions and of the $d=1$ dimension; namely, the spontaneously broken phase
should vanish for $d=1$ which indicates that the saddle point and the nontrivial 
IR fixed point ($1/\bar z_{\mr{IR}} \equiv 0$, $\bar u_{\mr {IR}} \equiv 1$) should
coincide. Thus, the distance between the nontrivial IR fixed point and the saddle 
point (see \fig{fig1}),
\bea
\label{dist}
D &\equiv& 
\sqrt{(\bar u_{\mr {IR}} - \bar u_{\star})^2 +  (1/\bar z_{\mr{IR}} - 1/\bar z_{\star})^2}
\nonumber \\
&=& \sqrt{(1 - \bar u_{\star})^2 +  1/\bar z_{\star}^2}
\eea
can be used to optimize the scheme dependence of RG equations;
i.e., the better the RG scheme the smaller the distance $D$ is. The other 
attractive IR fixed point ($\bar u_{k\to 0} = 0$, $1/\bar z_{k\to 0} =\infty$)
corresponds to the symmetric phase \cite{cg_sg,css}.

\section{Optimization of the power-law regulator}
\begin{figure}[ht] 
\begin{center} 
\epsfig{file=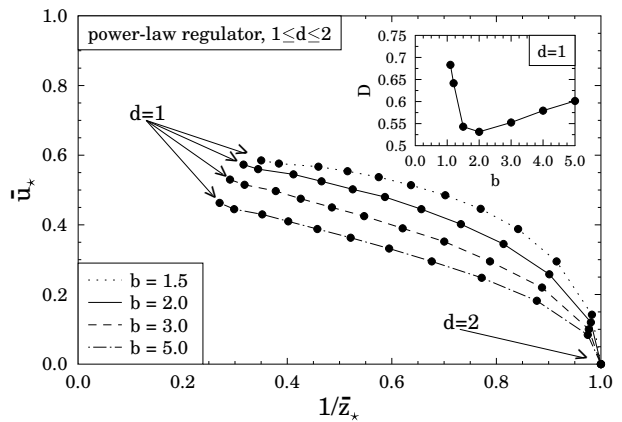,width=8.0 cm}
\caption{
\label{fig2}
Positions of the saddle point of the SG model for dimensions $1\leq d \leq 2$ 
obtained by the power-law regulator \eq{pow_lim}. The inset shows 
the dependence of the distance $D$ defined by Eq.\eq{dist} on the parameter 
$b$ for dimension $d=1$.
} 
\end{center}
\end{figure}

Let us consider first the optimization of RG equations using the power-law 
type regulator \eq{pow_lim} in the framework of the SG model. According to 
the numerical solution of Eqs. \eq{exact_u} and \eq{exact_z}, the 
saddle point appears in the RG flow for dimensions $1\leq d \leq 2$ and 
its position is plotted in \fig{fig2} for various values of the parameter 
$b$.
%
%
For $d=2$ dimensions the curves coincide since the fixed point at which 
the two-dimensional SG model undergoes a topological phase transition 
is at $\bar u_\star = 0$, $1/\bar z_{\star} =1$ scheme independently. For 
fractal dimensions and also for $d=1$ the position of the saddle point 
becomes scheme dependent, i.e., it depends on the parameter $b$ of the 
regulator function \eq{pow_lim}. Since the (spontaneously) broken phase 
should vanish for the $d=1$ dimension, the distance Eq.\eq{dist} between the 
saddle point and the nontrivial IR fixed point ($1/\bar z_{\mr{IR}} \equiv 0$, 
$\bar u_{\mr {IR}} \equiv 1$) can be used to optimize the RG equation. The 
inset shows the dependence of the distance $D$ on the parameter 
$b$ and indicates that $b=2$ is the optimal choice. Thus, it recovers known 
results obtained by optimization \cite{opt_func,opt_rg} based on the optimal 
convergence of the flow that validates the strategy proposed here.

\section{Optimization of the CSS regulator}
\begin{figure}[ht] 
\begin{center} 
\epsfig{file=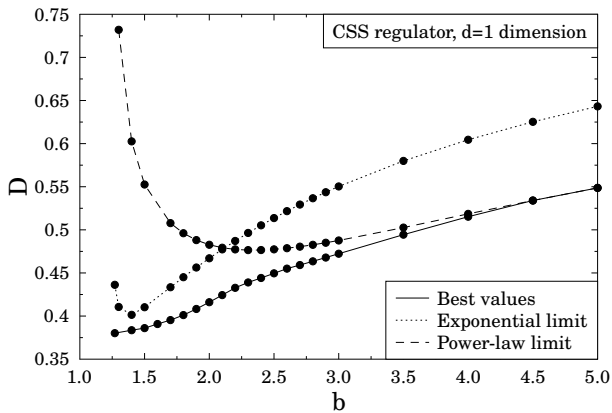,width=8.0 cm}
\caption{
\label{fig3}
Dependence of $D$ \eq{dist} on the parameter $b$ of the CSS regulator 
\eq{css_norm} represented in various limits.  Best values are
obtained in terms of the parameters $c$ and $h$, for "fixed" $b$.} 
\end{center}
\end{figure}

Let us perform the optimization of the normalized CSS regulator 
\eq{css_norm} in the framework of the one-dimensional SG model using 
the optimization scenario based on the minimization of the distance $D$
defined by Eq. \eq{dist}. First, one has to determine the position of the 
saddle point. This can be done by using the linearized form of RG equations 
\eq{exact_u}, and \eq{exact_z} with dimensionless variables (linearized in terms 
of $\tilde u$) since usually $\tilde u_\star$ is found to be much smaller than one 
(after that $\bar u_\star$, $1/\bar z_\star$ and the distance $D$ can be calculated). 
Let us first perform consistency checks. In \fig{fig3} we plot the dependence of 
$D$ on the parameter $b$ of the CSS regulator \eq{css_norm} in various 
limits. For example, findings determined by the power-law limit \eq{pow_lim} 
of the CSS regulator (with $c=0.0001$ and $h=0.0001$) is represented by 
the dashed line which can be compared to the inset of \fig{fig2} where results 
obtained by the exact flow equations \eq{exact_u}, and \eq{exact_z} are shown. 
The two curves are qualitatively the same, and both have a minima at 
$b\approx 2$ (in case of \fig{fig3} it is at $b = 2.3$). Let us remind the reader 
that $b=2$ is the optimal choice according to the Litim-Pawlowski method.
Another consistency check is based on the results obtained by the exponential 
limit \eq{exp_lim} of the CSS regulator (with $c=1$, $h=0.0001$) which is 
shown by the dotted line in \fig{fig3}. The minimum (i.e. the optimal choice)
is at $b\approx 1.4$ whereas the Litimi-Pawlowski optimization indicates 
$b=1.44$.
%
%
The full line of \fig{fig3} shows the minimum (i.e. the best) values of $D$ 
obtained in terms of the parameters $c$ and $h$, for "fixed" $b$. For large
$b$ it coincides with the power-law limit. It has an inflection point at 
$b \approx 2.1$ where the power-law and exponential limits cross each
other. For small $b$ the best values are obtained for small but nonzero $c$. It also 
clearly indicates that the most favorable choice is $b\approx 1$. Thus, the 
Litim limit ($b\approx 1$, $c\approx 0$) of the CSS regulator is found to be 
the optimal choice beyond LPA (here we use the Litim limit to refer to a small 
but nonzero value for $c$ and do not take $c\to 0$ exactly). The computation 
of the best value of $D$ for $b\to 1$ is costly since the derivatives of the CSS 
regulator for $c\to 0$ have an oscillatory behavior \cite{css}. Nevertheless for 
small but finite $c$ the derivatives always exist; hence the Litim limit of the 
CSS can always be used at any order of the gradient expansion. It does not 
hold for the Litim regulator itself, which confronts the gradient expansion. 
Another important observation is that the usage of the PMS method (the global 
extremum of the CSS) produces exactly the same optimal parameters. 
Similarly, in LPA \cite{css_pms} the Litim limit of the CSS with $h=1$ was 
found to be the most favorable regulator. Beyond LPA the optimal choice for 
$h$ could depend on the model and also the approximations used. For example, 
here we found $c = 0.1$ and $h\approx 0.3$ as the optimal parameters for 
$b=1.25$.

\section{Summary}
A new optimization procedure for the functional RG method has been 
discussed that is based on the requirement for the absence of spontaneous 
symmetry breaking in the $d=1$ dimension. It has been applied for the SG
model where no field-dependent wave-function renormalization is required; 
hence the method is suitable for optimization beyond LPA. It is validated by 
recovering known results on the power-law and exponential regulators. The 
CSS regulator has been optimized which leads to the best choice among 
the class of regulator functions. Results were obtained beyond LPA here and 
the Litim limit of the CSS was found to be 
the optimal choice. This is supported also by the PMS method that was tested 
in LPA \cite{css_pms} for the $O(N)$ model and for QED$_2$. Therefore, 
considerations were done for three different models, in three different dimensions 
at various orders of the derivative expansion with various optimization methods and
all these results indicate that the Litim limit of the CSS (small but nonzero $c$) 
is the most favorable choice.

\end{document}